\documentclass[twocolumn,showpacs,preprintnumbers,amsmath,amssymb,showkeys]{revtex4}

\usepackage{graphicx}
\usepackage{dcolumn}
\usepackage{bm}
\usepackage{bezier}

\begin{document}

\title{Lasing and transport in a coupled quantum dot-resonator system}

\author{Pei-Qing Jin}
\affiliation {Institut f\"ur Theoretische Festk\"orperphysik,
      Karlsruhe Institute of Technology, 76128 Karlsruhe, Germany}
\author{Michael Marthaler}
\affiliation {Institut f\"ur Theoretische Festk\"orperphysik,
      Karlsruhe Institute of Technology, 76128 Karlsruhe, Germany}
\author{Jared H.~Cole}
\affiliation {Chemical and Quantum Physics, School of Applied Sciences,
RMIT University, Melbourne 3001, Australia}
\author{Alexander Shnirman}
\affiliation{Institut f\"{u}r Theorie der Kondensierten Materie,
Karlsruhe Institute of Technology, 76128 Karlsruhe, Germany}
\affiliation {DFG Center for Functional Nanostructures (CFN),
      Karlsruhe Institute of Technology, 76128 Karlsruhe, Germany}
\author{Gerd Sch\"on}
\affiliation {Institut f\"ur Theoretische Festk\"orperphysik,
      Karlsruhe Institute of Technology, 76128 Karlsruhe, Germany}
\affiliation {DFG Center for Functional Nanostructures (CFN),
      Karlsruhe Institute of Technology, 76128 Karlsruhe, Germany}

\date{\today}

\begin{abstract}
We investigate a double quantum dot coupled to a transmission line resonator.
By driving a current through the double dot, 
a population inversion between the dot levels can be created, and
a lasing state of the radiation field is generated within a sharp resonance window.
The transport current correlates with the lasing state.
The sharp resonance condition allows for resolving small differences in the dot properties.
Dissipative processes in the quantum dots and their effect on the lasing and transport behavior
are investigated.
\end{abstract}
\maketitle

\section{Introduction}

The interaction between light and matter is one of the fundamental topics in physics.
Modern techniques allow for fabricating artificial atoms with highly controllable parameters,
which provide an ideal testing ground for quantum optics.
In the emerging field of
circuit quantum electrodynamics (QED) \cite{Wallraff04,Chiorescu04,Blais04,Schoelkopf,Marthaler11},
where superconducting qubits, playing the role of artificial atoms,
are strongly coupled to a superconducting transmission line resonator,
many quantum optics effects have been demonstrated with unprecedented accuracy,
and some novel phenomena have been observed.
An example is single-qubit lasing \cite{Astafiev,hauss08,grajcar},
where quantum noise influences the linewidth of the emission spectrum in a characteristic way
\cite{SQL1,SQL2,SQL3,Didier}.
Recently proposals to replace in circuit QED setups the
superconducting qubits by semiconductor quantum dots
were put forward \cite{Lukin04,Burkard06,Trif,Cottet,LWD}.
Quantum dots have the advantages that many of the parameters,
e.g., the level structure, coupling strength, and tunneling rates, can be varied {\sl in situ},
and details of the transport properties can be resolved via adjacent quantum point contacts \cite{Tarucha,Oosterkamp,Fujisawa}. 
Fast experimental progress has been made towards coupling quantum dots to a superconducting resonator \cite{DR1,DR2,DR3}.

In this paper we consider such a circuit QED setup
with a semiconductor double quantum dot coupled to a high-Q transmission line.
We find that by driving a current through a suitably biased double dot,
a population inversion can be created between the dot levels,
which then may lead to a lasing state in the resonator.
This lasing state in the resonator correlates with the transport properties through the double dot system.
This property not only allows probing the lasing state via a current measurement,
which may be easier to perform in an experiment,
but also, because of the  narrow resonance window,
opens perspectives for applications for high resolution measurements.

The paper is organized as follows.
In Sec.~\ref{Sec:MM} we present the model for the coupled quantum dot-resonator system
and the pumping mechanism,  provide estimates for the parameters, and we introduce
the theoretical tools for the analysis of the system. 
In Sec.~\ref{Sec:SP} we investigate the stationary properties of the radiation field inside the resonator,
while its emission spectrum and linewidth are discussed in Sec.~\ref{Sec:LW}.
In Sec.~\ref{Sec:CLT} we analyze the correlation between the lasing and transport properties
of this coupled system.
We conclude with a brief summary.

\section{Model and method}\label{Sec:MM}

\begin{figure}[t]
\centering
\includegraphics[width=0.3\textwidth]{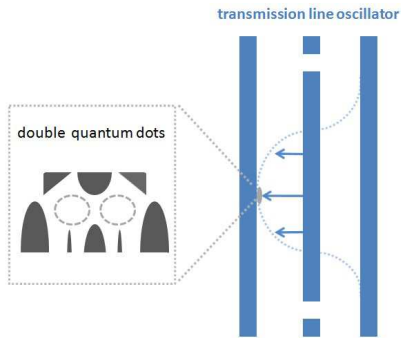}
\caption{(Color online) Coupled quantum dot-resonator circuit.
The double dot is placed at a maximum of the electric field of the transmission line
in order to maximize the  dipole interaction with the resonator.
For gate-defined quantum dots most parameters can be tuned by the applied voltages. }\label{fig:dot}
\end{figure}

We consider a system shown schematically in Fig.~\ref{fig:dot}.
The double quantum dot is biased such that three charge states play a role,
a reference neutral state and two states differing by a single extra electron 
in the left or right dot. 
The three states are denoted by $|0,0\rangle$,
$|1,0\rangle$, and $|0,1\rangle$, respectively. 
The two singly charged states, with energy difference $\epsilon$ (denoted as dot detuning),
are coupled by coherent interdot tunneling with strength $t$.
The double dot system is therefore described by a two-level Hamiltonian,
\begin{eqnarray}\label{eq:DD}
 H_{\rm dd} = \frac{1}{2} (\epsilon\, \tau_z + t\, \tau_x),
\end{eqnarray}
with Pauli matrices $\tau_z = |1,0\rangle\langle 1,0|-|0,1\rangle\langle 0,1|$
and $\tau_x= |1,0\rangle\langle 0,1|+|0,1\rangle\langle 1,0|$.

We model the superconducting transmission line 
as a harmonic oscillator with frequency $\omega_{\rm r}$,
and $a^{}$ and $a^\dag$ denoting the annihilation  and creation operators
for the radiation field.
In typical circuit QED experiments, the resonator frequency is of the order of several GHz,
and the vacuum-fluctuations-induced voltage drop $V_{\rm r}$ 
between its central strip and ground plane is of the order of $\mu$V  \cite{Wallraff04}.
For definiteness we choose $\omega_{\rm r}/2\pi=4\, {\rm GHz}$ throughout this paper.
For the geometry shown in Fig.~\ref{fig:dot},
the superconducting resonator is capacitively coupled to
the charge states of the double dot which have different dipole moments.
It leads to the interaction 
\begin{eqnarray}
 H_{\rm c} = \hbar\, g_0\, (a^\dag+a^{}) \tau_z \, . 
\end{eqnarray}
When the interdot tunneling is weak and
the single-particle wavefunctions are strongly localized in either dot,
the coupling strength can be estimated as $g_0 \sim eEd/(2\hbar) \sim 50~{\rm MHz}$,
where $d \sim 0.3~\rm{\mu m}$ is the distance between the centers of the two dots
and $E \sim 0.2~\rm{V}/\rm{m}$ the electric field at the antinode of the resonator mode.
This estimate agrees with the experimental observations~\cite{Wallraff2012}.

In the eigenbasis of the two-level Hamiltonian (\ref{eq:DD}),
within the rotating wave approximation,
the Hamiltonian for the coupled system reduces to
\begin{eqnarray}\label{eq:ChargeH}
H_{\rm{sys}}  = \frac{\hbar\omega_0}{2} \sigma_z
 + \hbar \omega^{}_{\rm r} a^\dag a
 + \hbar g (a^\dag \sigma_- +a^{} \sigma_+).
\end{eqnarray}
Here $\sigma_z = |e\rangle\langle e| - |g\rangle\langle g|$
(similar for $\sigma_\pm$)
is defined in this subspace spanned by eigenstates
\begin{eqnarray}
 |e\rangle &=& \cos\left(\theta/2\right)|1,0\rangle + \sin\left(\theta/2\right)|0,1\rangle,
               \nonumber \\
 |g\rangle &=& -\sin\left(\theta/2\right)|1,0\rangle + \cos\left(\theta/2\right)|0,1\rangle,
\end{eqnarray}
with angle $\theta = \arctan(t/\epsilon)$ characterizing the mixture of the charge states.
The energy splitting between the eigenstates, $\hbar \, \omega_0 = \sqrt{\epsilon^2+t^2}$,
can be tuned via gate voltages,
allowing control of the detuning relative to
 the resonator frequency $\Delta = \omega_0-\omega_{\rm r}$.
The effective dot-resonator coupling strength, $g = g^{}_0 \sin\theta$,
reaches its maximum at $\epsilon=0$.
To achieve a lasing state, the detuning $\Delta$ should be small
and the coupling $g$ strong enough to overcome the dissipative processes.
This requires both the energy difference $\epsilon$
and the interdot tunneling $t$ to be of the order of several $\mu{\rm eV}$.

\begin{figure}[t]
\centering
\includegraphics[width=0.35\textwidth]{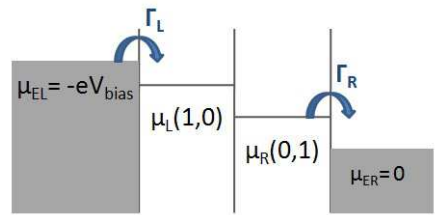}
\caption{(Color online) Tunneling sequence in the double dot system.
The chemical potentials $\mu_{\rm L} (1,0)$ and $\mu_{\rm R} (0,1)$ of the two dots
are assumed to be arranged as indicated.}\label{fig:pump}
\end{figure}
In order to create a population inversion, which is usually required for lasing behavior,
we consider a pumping scheme involving the third state $|0,0\rangle$,
similar to an optical laser.
(Note that lasing without inversion is also possible in circuit QED setups
where the dissipative environment enhances the 
photon emission as compared to absorption \cite{LWI}.
Population inversion induced by other pumping mechanisms,
e.g., driving the two-level system coherently \cite{hauss08,Carmichael2011},
were also discussed.)
As indicated in Fig. \ref{fig:pump}, we assume that a bias voltage
is applied across the double dot such that
only the chemical potentials of the states $|1,0\rangle$ and $|0,1\rangle$
lie within the bias window.
At low temperature compared to the charging energy,
the only possibility for an electron to tunnel into the dot system
is from the left lead to the left dot,
leading to the transition from state $|0,0\rangle$ to $|1,0\rangle$
with tunneling rate $\Gamma_{\rm L}$.
Similarly, an electron can tunnel out into the right lead,
creating a transition from $|0,1\rangle$ to $|0,0\rangle$
with tunneling rate $\Gamma_{\rm R}$.
The pumping leads to a non-equilibrium state
where the population in the state $|1,0\rangle$ with higher energy is enhanced
compared to that in $|0,1\rangle$.
The population inversion persists when we go to the eigenbasis, where
it is given by $ \tau_0  = 4 \cos\theta / [3+\cos(2\theta) ]$ \cite{LWD}.

To analyze the dynamics of the coupled dot-resonator system,
we employ a master equation for the reduced density matrix $\rho$
in the Born-Markovian approximation,
\begin{eqnarray} \label{eq:ME}
 \dot \rho &=& -\frac{i}{\hbar}\left[H_{\rm_{sys}}, \rho\right]
  +\mathcal L_{\rm r}\, \rho + \mathcal L_{\rm \downarrow}\rho+\mathcal L_{\rm \varphi}\, \rho
  + \mathcal L_{\rm L}\, \rho+\mathcal L_{\rm R}\, \rho
                            \nonumber \\
 &\equiv& \mathcal L_{\rm tot} \, \rho \, .
\end{eqnarray}
We assume that the dissipative environment is characterized by
 smooth spectral functions \cite{Gardiner,Carmichael} and can be described by 
 Lindblad operators of the form
\begin{equation}
 {\cal L}_{i}\rho=\frac{\Gamma_i}{2}\left(2L_i\rho L_i^{\dag}-L_i^{\dag}L_i\rho-\rho L_i^{\dag}L_i\right).
\end{equation}
For the oscillator we take the standard decay terms $L_{\rm r}=a$ with rate $\Gamma_{\rm r}=\kappa$.
Throughout this paper we consider low temperatures, $T = 0$,
with vanishing thermal photon number and excitation rates.
For the two-level system we account for the relaxation by $L_\downarrow =\sigma_-$ with rate $\Gamma_\downarrow$
and for pure dephasing by $L_{\varphi}=\sigma_z$ with rate $\Gamma_\varphi^*$.
For the discussion of lasing behavior,
it is worthwhile to introduce a total decoherence rate for the dot system, namely,
$\Gamma_\varphi =\Gamma_\downarrow/2 + \Gamma\Bigl[\sin^4(\theta/2)+\cos^4(\theta/2)\Bigr]/2 +\Gamma_\varphi^*$.
The last two terms account for the incoherent tunneling between the
electrodes and the left and right dots.
In the basis of the pure charge states, the Lindblad operators describing these processes
are given by $ \mathcal L_{\rm L} = |1,0 \rangle\langle 0,0|$ 
and $ \mathcal L_{\rm R} = |0,0 \rangle\langle 0,1|$.
For simplicity, we set $\Gamma_{\rm L} = \Gamma_{\rm R}=\Gamma$ throughout the paper.

\section{Stationary properties}\label{Sec:SP}

We start with analyzing the stationary properties of the radiation field inside the resonator,
namely the average photon number $\langle n\rangle$ and the 
Fano factor $F = (\langle n^2\rangle-\langle n\rangle^2)/\langle n\rangle$,
which serve as good indicators of the state of the radiation field.
\begin{figure}[t]
\centering
\includegraphics[width=0.53\textwidth]{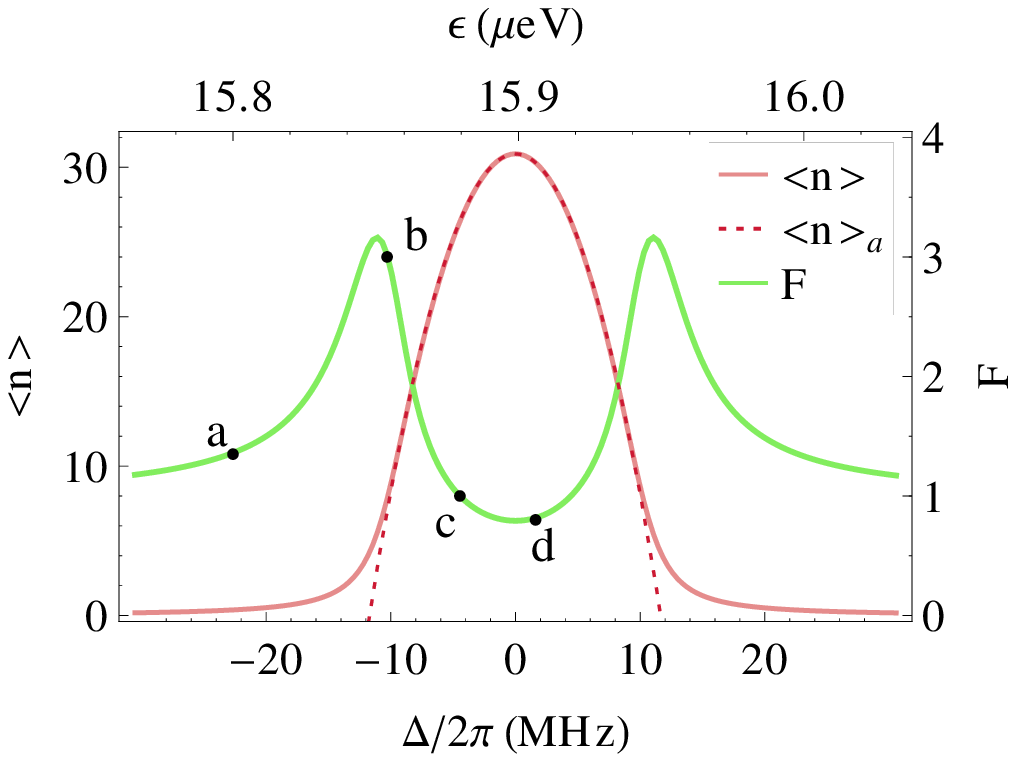}
\includegraphics[width=0.4\textwidth]{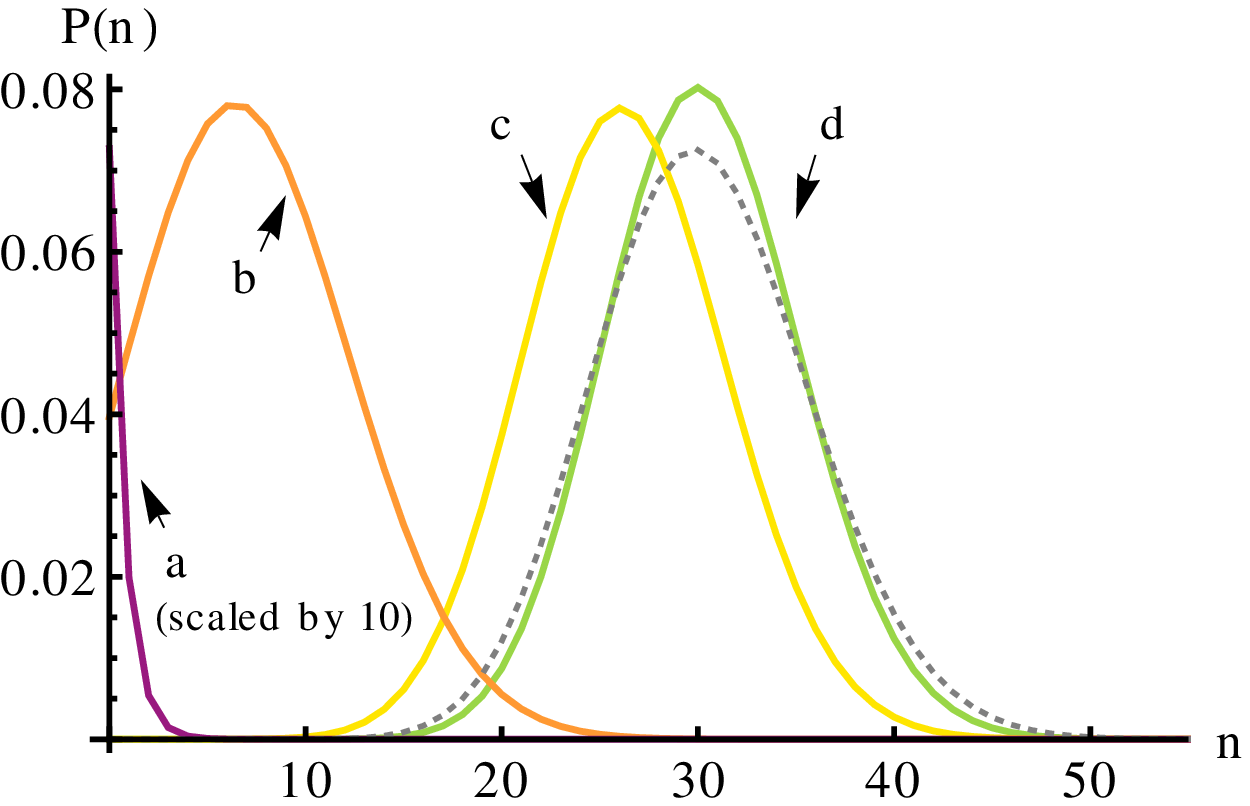}
\caption{
(Color online)
Upper panel: Average photon number $\langle n\rangle$
and Fano factor $F$ as functions of dot detuning $\epsilon$
for vanishing decoherence of the dot levels ($\Gamma_\downarrow = \Gamma_\varphi^* =0$).
The analytical approximation $\langle n\rangle_{\rm a}$ (dotted line) in Eq. (\ref{eq_Approximate_result_for_n})
is compared to the numerical result (solid line).
The decay rate of the resonator is chosen to be $\kappa/2\pi = 40 \, {\rm kHz}$.
Throughout this paper we choose the incoherent tunneling rate $\Gamma = 4 {\rm MHz}$,
the coherent interdot tunneling strength $t= 5\, {\rm \mu eV}$,
and the bare coupling strength $g_0 = 4 {\rm MHz}$.
Lower panel: Photon number distribution at points a, b, c, and d marked in the upper panel.
Dotted line indicates a Poissonian distribution for comparison with the sub-Poissonian distribution at point d.
}\label{fig:SP}
\end{figure}
We plot both 
as functions of the detuning in Fig.~\ref{fig:SP}.
To begin with we neglect the relaxation and decoherence of the dot levels, $\Gamma_\downarrow = \Gamma_\varphi^* =0$.
When the detuning is large (e.g., the point/curve marked by `a'  in Fig. \ref{fig:SP}),
the quantum dot effectively does not interact with the resonator,
and the photon number is low with a thermal distribution.
For weaker detuning (e.g., point `b') the photon number increases sharply and
saturates at resonance, $\Delta = 0$.
Approximately, the photon number can be estimated as \cite{Marthaler09},
\begin{eqnarray}\label{eq_Approximate_result_for_n}
 \langle n\rangle_{\rm a} = \frac{\Gamma \cos\theta}{3\, \kappa}
 -\frac{\cos(2\theta)+7 }{96\, g^2} (4\Delta^2+\Gamma^2).
\end{eqnarray}
In the typical parameter regime of experiments
the peak value of the photon number is mainly given by the first term,
i.e., the ratio between the incoherent pumping and the decay rate of the resonator.
For large photon number good agreement is reached
between the analytical expression (dotted line) and numerical results (solid line).

Statistical properties of the radiation field can be obtained by investigating the Fano factor.
When the resonator is in a lasing state (e.g., point `c'), 
the radiation field is in the coherent state and the photon number distribution is Poissonian.
In this case the Fano factor equals to 1.
In the strong-coupling regime, the Fano factor can even become smaller than 1 (e.g., point `d'),
which indicates a sub-Poissonian distribution of the radiation field. 
In this non-classical regime, the photon number distribution is squeezed compared to the Poissonian
distribution.

\begin{figure}[t]
\hspace{5mm}
\hspace{8mm} \includegraphics[width=0.44\textwidth]{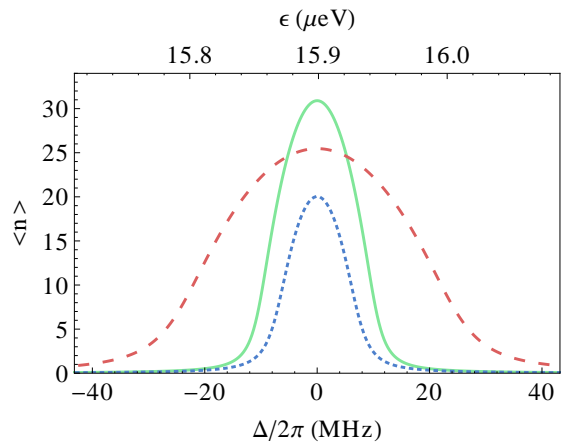} \\[6mm]
\includegraphics[width=0.44\textwidth]{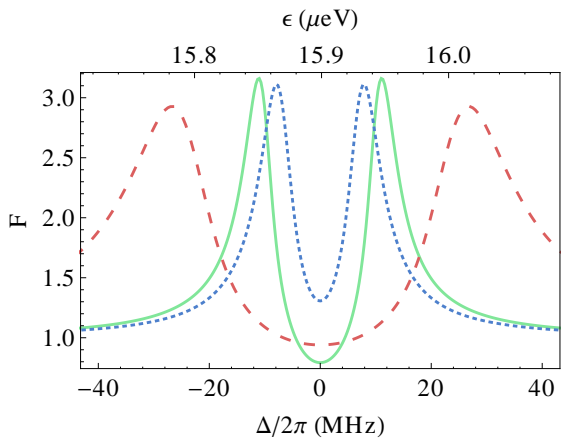}
\caption{(Color online)
Average photon number $\langle n \rangle$ and Fano factor $F$
as functions of detuning.
Solid lines represent the results without decoherence,
blue dotted lines are those for relaxation rate $\Gamma_\downarrow/2\pi = 1.2~ {\rm MHz}$
and vanishing pure dephasing,
while red dashed lines are for pure dephasing rate $\Gamma_\varphi^*/2\pi = 12~ {\rm MHz}$
but vanishing relaxation.
Other parameters are chosen to be the same in Fig. \ref{fig:SP}.
}\label{fig:DecCharge}
\end{figure}
We now turn to the effects of decoherence of the dot levels on the lasing state,
which are illustrated in Figs.~\ref{fig:DecCharge}.
The relaxation process reduces the pumping efficiency
and deteriorates the lasing state.
Hence the photon number decreases.
At the same time the Fano factor around resonance becomes larger than 1,
resulting from an increase of amplitude fluctuations.
Pure dephasing, on one hand, diminishes the efficiency of
energy exchange between the dot levels and the resonator.
On the other hand, it effectively broadens the window in
which the dot levels can interact with the resonator.
As a result, the peak of the photon number is broadened,
while the peak value decreases.

\section{Emission spectrum and linewidth}\label{Sec:LW}

\begin{figure}[t]
\centering
\includegraphics[width=0.43\textwidth]{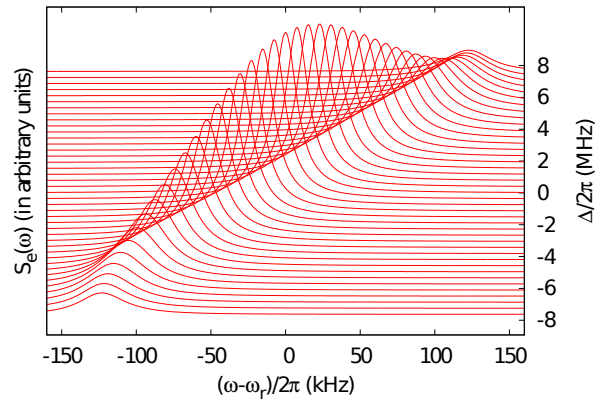}
\caption{
(Color online)
Emission spectrum $S_{\rm e}(\omega)$ (in arbitrary units)
for different detunings.
 }\label{fig:Spectrum}
\end{figure}
We now turn to the spectral properties of the radiation field.
The emission spectrum of the radiation field,
\begin{eqnarray}
 S_e(\omega) = \int^\infty_{-\infty} e^{-i\omega t}
 \langle a^\dag (t) a(0) \rangle,
\end{eqnarray}
which can be measured directly in experiments,
provides further insights into the state of the radiation field.
To calculate the phase correlator, we make use of the quantum regression theorem \cite{Gardiner},
\begin{eqnarray}
   \langle a^\dag (t) a(0) \rangle = {\rm Tr}[a^\dag \, e^{\mathcal L_{\rm tot} t} a \rho_{\rm st} ],
\end{eqnarray}
with $\rho_{\rm st}$ being the stationary solution of the master equation (\ref{eq:ME}).

The spectral properties of the radiation field in the quantum dot-resonator system
are similar to those of a single-qubit maser discussed in Refs. \cite{SQL1,SQL2,SQL3,Didier}.
In Fig.~\ref{fig:Spectrum} we plot the emission spectrum for different values of 
the detuning.
At resonance, $\Delta = 0$, the emission spectrum shows a sharp peak
at the resonator frequency $\omega_{\rm r}$.
For nonvanishing detuning the peak of the emission spectrum is shifted in frequency
by $\delta \omega$.
For weak detuning, this shift grows linearly with the detuning,
 $\delta \omega \simeq \Delta \kappa / 2\Gamma_\varphi$,
with $\Gamma_\varphi = \Gamma_\downarrow/2+\Gamma/4+\Gamma_\varphi^*$ characterizing the total dissipation rate
of the quantum dot system.
At the same time, the peak value of the emission spectrum decreases with growing detuning,
related to a lowering of the average photon number.

\begin{figure}[t]
\centering
\includegraphics[width=0.43\textwidth]{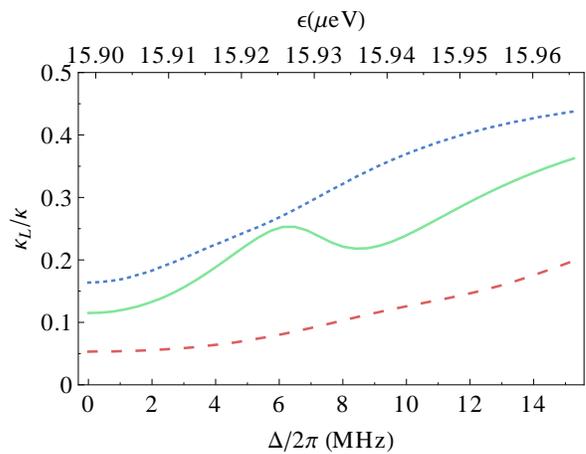}
\caption{
(Color online) Linewidth of the emission spectrum as a function of detuning.
The green solid line is the result for vanishing relaxation and decoherence
$\Gamma_\downarrow = \Gamma_\varphi^* = 0$, the blue dotted line for relaxation rate
$\Gamma_\downarrow/2\pi = 1.2\, {\rm MHz}$ but vanishing pure dephasing,
and the red dashed line for pure dephasing rate $\Gamma_\varphi^*/2\pi = 4\, {\rm MHz}$
but $\Gamma_\downarrow =  0$.
Here the decay rate of the resonator is chosen to be $\kappa = 80\, {\rm kHz}$.
}\label{fig:LW}
\end{figure}

The linewidth of the emission spectrum is plotted as a function of detuning in Fig.~\ref{fig:LW}.
In a typical situation where the coupling $g$ is not too strong compared to the decoherence 
of the coupled system
$(\kappa, \Gamma_\varphi)$,
the linewidth grows monotonously with increasing detuning, as indicated, e.g., by the blue dotted line.
At resonance,  deep in the lasing regime,
the linewidth can be estimated as $\kappa_{\rm L} \propto g^2/ (\Gamma_\varphi \langle n \rangle)$ \cite{SQL3}. It becomes very small for  large photon numbers.
Here, due to the restrictions of our numerical calculation, 
we illustrate to a situation with low photon number around 15 and not very small values of  
$\kappa_{\rm L}$.
For stronger detuning the system leaves the lasing regime,
and the linewidth is approximately given by
$\kappa_{\rm L} \simeq \kappa/2 -g^2\, \Gamma_\varphi\, \tau_0/ (\Gamma_\varphi^2 + \Delta^2)$,
which approaches the bare linewidth of the resonator, $\kappa/2$, for strong detuning.

When the coupling is much stronger than the decoherence rate,
the linewidth depends nomonotonously on the detuning
exhibiting an enhancement at the lasing transition (green solid line)  \cite{SQL3}.
Different from the conventional lasing situation
where the emission spectrum is mainly broadened by phase fluctuations,
in the strong coupling regime at the lasing transition,
the contribution from amplitude fluctuations is magnified
via its coupling to the phase fluctuations, leading to the peak shown in the figure. 
In this transition regime, the phase correlator $\langle a^\dag (t) a (0) \rangle$
exhibits a non-exponential decay in time.

A further surprising phenomenon occurs for a 
strong pure dephasing rate $\Gamma_\varphi^*/2\pi = 4 \, {\rm MHz}$.
Instead of broadening the emission spectrum, the pure dephasing actually reduces the linewidth (red dashed line),
leading to a much sharper spectrum peak.
An intuitive interpretation can be obtained by examining the linewidth at resonance,
where a semi-quantum approach is accessible \cite{SQL2}.
In this case, the linewidth at resonance in the deep lasing regime
can be estimated as $\kappa_{\rm L} \propto g^2/ (\Gamma_\varphi \langle n \rangle) $.
With growing pure dephasing, the photon number changes little
while the total decoherence rate $\Gamma_\varphi$ increases linearly with the pure dephasing rate,
which leads to the decrease in the linewidth.

\section{Correlations between lasing and transport properties}\label{Sec:CLT}

\begin{figure}[t]
\centering
\includegraphics[width=0.4\textwidth]{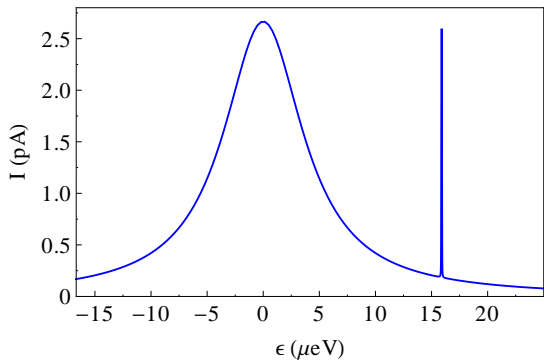}
\caption{(Color online) Current as a function of dot detuning $\epsilon$.
The parameters are chosen to be the same as in Fig. \ref{fig:SP}.
The dot levels become resonant with the resonator at $15.9\, {\rm \mu eV}$.
}\label{fig:curr}
\end{figure}
\begin{figure}[t]
\centering
\includegraphics[width=0.42\textwidth]{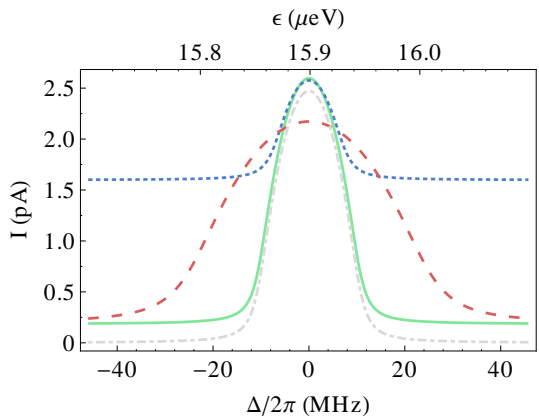}
\caption{(Color online) Current as a function of detuning $\Delta$
with the green solid line for the result with vanishing decoherence
$\Gamma_\downarrow = \Gamma_\varphi^* = 0$, the blue dotted line with relaxation rate
$\Gamma_\downarrow/2\pi = 1.2\, {\rm MHz}$ but vanishing pure dephasing,
and the red dashed line with pure dephasing rate $\Gamma_\varphi^*/2\pi = 4\, {\rm MHz}$
but $\Gamma_\downarrow =  0$.
The photon number (gray dot-dashed line) is also shown for comparison. 
}\label{fig:CurrDec}
\end{figure}

A current flows in the system if the tunneling cycle
$|0,0\rangle\rightarrow|1,0\rangle\rightarrow|0,1\rangle\rightarrow|0,0\rangle$ is completed.
This can be achieved by either coherent interdot tunneling or a coherent process involving the 
excition of photons in the resonator.
We calculate the current using
\begin{eqnarray}\label{eq:curr}
 I = e\sum_{i,j}\Gamma_{i\rightarrow j}\,\langle i|\rho_{\rm st} |i\rangle,
\end{eqnarray}
where the index $i$ refers to the states $|g\rangle$, $|e\rangle$,
and $|0,0\rangle$, and $\Gamma_{i\rightarrow j}$ denotes the
transition rate from state $|i\rangle$ to $|j\rangle$.
In Fig.~\ref{fig:curr} we plot the current as a function of the dot detuning $\epsilon$.
As expected, two resonance peaks show up in the current.
The coherent interdot tunneling leads to the broad peak around $\epsilon=0$
with width given by the tunneling rate $t$ (here $t\gg \hbar\,\Gamma$)~\cite{Vaart,Stoof}.
Interestingly, the second peak due to transitions involving the excitation of photons is -- for realistic values of the parameters -- much narrower.
As shown in Fig. \ref{fig:CurrDec}, it correlates with the lasing state, since the excitation of a photon
in the resonator is caused by the electron tunneling between the two dots.
Both the lasing state and the current peak exist only in a narrow ``resonance window" $|\Delta|\le W/2$.
This window can be estimated from Eq. (\ref{eq_Approximate_result_for_n}) with condition $\langle n \rangle_{\rm a} \ge 0$,
namely,
\begin{eqnarray}
 W = \Gamma \sqrt{\frac{32 \cos\theta\, g^2}{\left[\cos(2\theta)+7\right]\kappa\,\Gamma}-1},
\end{eqnarray}
which reduces to $W\approx 2 g\sqrt{\Gamma/\kappa-1}$ for small $\theta$.
To estimate the height of the narrow current peak,
we adopt an adiabatic approximation assuming the dynamics of the resonator
to be much slower than that of the quantum dots \cite{Scully}.
For $\kappa\ll \Gamma$ and small $\theta$, 
the peak value of the current is given by
\begin{eqnarray}
 I(\Delta=0) \simeq e \Gamma \sum_{n=0}^{\infty} P(n)
 \left[\frac{2(n+1)}{3(n+1)+\Gamma^2/(4g^2)} \right].
\end{eqnarray}
Here $P(n)\simeq (\Gamma/\kappa) P(0)\Pi_{l=1}^{n} [3l+\Gamma^2/(4 g^2)]^{-1}$
denotes the probability of having $n$ photons in the resonator.
When the coupling to the resonator is strong compared to the incoherent tunneling $\Gamma$,
the peak current approaches $2e\Gamma/3$.

The correlation between the lasing state and the transport current is remarkable in two ways.
On one hand, the current peak, which may be easier to measure than the photon  state of the
resonator, can be used as a probe of the lasing state. On the other hand, the rather
sharp resonance condition needed for the lasing makes the current peak narrow,
while at the same time the value of the current is reasonably high.
This allows resolving in an experiment small details of the dot properties.

The effects of decoherence on the transport current are illustrated in Fig. \ref{fig:CurrDec}.
Relaxation, on one hand, reduces the efficiency of the coherent transition between the excited and ground states.
Hence within the resonance window the current induced by the coherent transition decreases.
On the other hand, relaxation also opens up an incoherent channel which increases the  current.
Pure dephasing leads to a broadening of the resonance peak in the current, as well as of the photon number.

\section{Summary}\label{Sec:Summary}

We have investigated a double quantum dot coherently coupled to a transmission line resonator.
A population inversion between the dot levels is created by a
pumping process due to a voltage applied across the double dot.
A lasing state of the radiation field develops within a sharp resonance window.
It correlates with a peak in the transport current.
The sharp resonance condition allows for resolving small differences in the dot properties.
This opens perspectives for applications of the setup and operation principle for high resolution measurements.

We analyzed the effects of dissipation on the dot levels.
Both relaxation and pure dephasing deteriorate the lasing state,
accompanied by a decrease in the photon number and an increase of amplitude fluctuations.
Relaxation processes shorten the life time of the two-level system and
reduce the population inversion and hence the lasing effect.
Pure dephasing, which also shortens the life time,
broadens the effective resonance window between the quantum dots and resonator.
Surprisingly, the emission spectrum can become even sharper with moderate
pure dephasing rate.
For the transport current, the relaxation opens up an extra incoherent channel
which increases the current outside the resonance window with the resonator.

\begin{acknowledgments}

We acknowledge fruitful discussions with S. Andr\'{e}, A. Romito and J. Weis,
as well as the support from the Baden-W\"{u}rttemberg Stiftung
via the 'Kompetenznetz Funktionelle Nanostrukturen' and the DFG via the 
Priority Program 'Semiconductor Spintronics'.

\end{acknowledgments}


\begin{thebibliography}{10}

\bibitem{Wallraff04}
A. Wallraff \emph{et al.}, Nature {\bf 431}, 162 (2004).

\bibitem{Chiorescu04}
I. Chiorescu \emph{et al.}, Nature  {\bf 431}, 159 (2004).

\bibitem{Blais04}
A. Blais \emph{et al.},  Phys. Rev. A {\bf 69}, 062320 (2004).

\bibitem{Schoelkopf}
R. Schoelkopf and S. Girvin, Nature \textbf{451}, 664 (2008).


\bibitem{Marthaler11}
M. Marthaler, J. Lepp\"{a}kangas, and J. H. Cole,
Phys. Rev. B {\bf 83}, 180505(R) (2011).


\bibitem{Astafiev}
O. Astafiev \emph{et al.}, Nature \textbf{449}, 588 (2007).

\bibitem{hauss08}
J. Hauss, A. Fedorov, C. Hutter, A. Shnirman, and G. Sch\"on, Phys. Rev. Lett. \textbf{100}, 037003
(2008).

\bibitem{grajcar}
M. Grajcar \emph{et al.}, Nature Physics \textbf{4}, 612 (2008).

\bibitem{SQL1}
S. Andr\'e, V. Brosco, A. Shnirman, and G. Sch\"on,
Phys. Rev. A \textbf{79}, 053848 (2009).

\bibitem{SQL2}
S. Andr\'e, V. Brosco, M. Marthaler, A. Shnirman, and G. Sch\"on,
Physica Scripta \textbf{T137}, 014016 (2009).

\bibitem{SQL3}
S. Andr\'e, P. Q. Jin, V. Brosco, J. H. Cole, A. Romito, A. Shnirman, and G. Sch\"on
Phys. Rev. A {\bf 82}, 053802 (2010).

\bibitem{Didier}
N. Didier, Ya. M. Blanter, and F. W. J. Hekking, Phys. Rev. B {\bf
82}, 214507 (2010).



\bibitem{Lukin04}
L. Childress, A. S. S\o{}rensen, and M. D. Lukin, Phys. Rev. A {\bf 69}, 042302 (2004).


\bibitem{Burkard06}
G. Burkard and A. Imamo\={g}lu, Phys. Rev. B {\bf 74}, 041307 (2006).


\bibitem{Trif}
M. Trif, V. N. Golovach, and D. Loss, Phys. Rev. B {\bf 77}, 045434
(2008).


\bibitem{Cottet}
A. Cottet and T. Kontos, Phys. Rev. Lett. {\bf 105}, 160502 (2010).


\bibitem{LWD}
P.-Q. Jin, M. Marthaler, J. H. Cole, A. Shnirman, and G. Sch\"{o}n,
Phys. Rev. B {\bf 84}, 035322 (2011).


\bibitem{Tarucha}
S. Tarucha \emph{et al.}, Phys. Rev. Lett. {\bf 77}, 3613 (1996).

\bibitem{Oosterkamp}
T. H. Oosterkamp \emph{et al.}, Nature {\bf 395}, 873 (1998).

\bibitem{Fujisawa}
T. Fujisawa \emph{et al.}, Science {\bf 282}, 932 (1998).

\bibitem{DR1}
T. Frey \emph{et al.}, Appl. Phys. Lett. {\bf 98}, 262105 (2011).


\bibitem{DR2}
T. Frey \emph{et al.}, Phys. Rev. Lett. {\bf 108}, 046807 (2012).


\bibitem{DR3}
M.R. Delbecq \emph{et al.}, Phys. Rev. Lett. {\bf 107}, 256804 (2011).


\bibitem{Wallraff2012}
T. Frey, P. J. Leek, M. Beck, A. Blais, T. Ihn, K. Ensslin, and A. Wallraff,
Phys. Rev. Lett. {\bf 108}, 046807 (2012).

\bibitem{LWI}
M. Marthaler, Y. Utsumi, D. S. Golubev, A. Shnirman, and Gerd Sch\"{o}n,
Phys. Rev. Lett. {\bf 107}, 093901 (2011).


\bibitem{Carmichael2011}
S. Hughes, and H. J. Carmichael,
Phys. Rev. Lett. {\bf 107}, 193601 (2011).




\bibitem{Gardiner}
C. W. Gardiner and P. Zoller, \emph{Quantum Noise}, (Springer, Berlin, 2004).


\bibitem{Carmichael}
H. J. Carmichael, \emph{Statistical Methods in Quantum Optics 1}, (Springer, Berlin, 2002).

\bibitem{Marthaler09}
M. Marthaler, PhD thesis, University Karlsruhe (2009).


\bibitem{Vaart}
N. C. van der Vaart \emph{et al.}, Phys. Rev. Lett. {\bf 74}, 4702 (1995).


\bibitem{Stoof}
T.H. Stoof and Yu.V. Nazarov, Phys. Rev. B {\bf 53}, 1050 (1996).


\bibitem{Scully}
M. O. Scully and M. S. Zubairy, \textit{Quantum Optics}, (Cambridge Press, Cambrige, 1997).














\end{thebibliography}
\end{document}